\documentclass[sigconf,screen,nonacm,review=false,timestamp=false]{acmart}
\AtBeginDocument{%
  }

\usepackage{colortbl,xcolor} 
\usepackage{arydshln} 
\usepackage{subcaption} 
\usepackage{listings}
\usepackage{comment}
\usepackage{amsmath}
\usepackage{multirow}
\usepackage{booktabs}
\usepackage{makecell}

\lstdefinelanguage{cypher}{
  morekeywords={MATCH, WHERE, RETURN, AND, OR, NOT, AS},
  sensitive=true,
  morecomment=[l]{//},
  morestring=[b]"
}

\begin{document}

\title{Ask Safely: Privacy-Aware LLM Query Generation for Knowledge Graphs*}

\author{Mauro Dalle Lucca Tosi}
\email{mauro.dalle-lucca-tosi@list.lu}
\orcid{0000-0002-0218-2413}
\affiliation{%
  \institution{Luxembourg Institute of Science and Technology (LIST)}
  \city{Esch-sur-Alzette}
  \country{Luxembourg}
}

\author{Jordi Cabot}
\email{jordi.cabot@list.lu}
\orcid{0000-0003-2418-2489}
\affiliation{%
  \institution{Luxembourg Institute of Science and Technology (LIST)}
}
\affiliation{%
  \institution{University of Luxembourg (UNILU)}
  \city{Esch-sur-Alzette}
  \country{Luxembourg}
}

\renewcommand{\shortauthors}{Tosi and Cabot}

\begin{abstract}
  Large Language Models (LLMs) are increasingly used to query knowledge graphs (KGs) due to their strong semantic understanding and extrapolation capabilities compared to traditional approaches. However, when KGs contain sensitive information and users lack local access to generative models, privacy becomes a critical concern. To address this issue, we propose a privacy-aware query generation approach for KGs. Our method identifies sensitive information in the graph based on its structure and omits such values before requesting the LLM to translate natural language questions into Cypher queries. Experimental results show that our approach effectively prevents sensitive data from being transmitted to third‑party services, while maintaining a high level of query accuracy.
\end{abstract}



\maketitle
\renewcommand{\thefootnote}{\fnsymbol{footnote}}
\footnotetext[1]{This version of the article has been accepted for publication at the 20th International Conference on Research Challenges
in Information Science (RCIS 2026), after peer review and is subject to Springer Nature’s AM terms of use, but is not the Version of Record and does not reflect post-acceptance improvements, or any corrections. The Version of Record is available online at: \url{https://doi.org/10.1007/978-3-032-26836-5_4}}
\renewcommand{\thefootnote}{\arabic{footnote}} 

\section{Introduction}

Large Language Models (LLMs) are increasingly used to retrieve knowledge. However, their responses may be inaccurate or contain hallucinations. To mitigate this, many applications rely on Retrieval-Augmented Generation (RAG), especially when retrieving knowledge from documents. Extending RAG‑style retrieval to Knowledge Graphs (KGs) requires providing the model with contextual graph data --- a process that can compromise privacy when the KG or the user’s query includes sensitive information~\cite{10.1145/3703155}.

This poses challenges when users want to query KGs but lack the resources to deploy their own generative models, and when the question, the KG, or both involve sensitive data. Such situations are common in practice, for example in healthcare~\cite{May03042022}, biomedical~\cite{10.1093/bioinformatics/btae353}, or legal~\cite{LAI2024181} use cases, where privacy is critical. In these cases, existing approaches that transmit graph data to external LLMs are not satisfactory.

Before generative LLMs, several rule-based~\cite{meloni2023integrating} and classical Machine Learning (ML)~\cite{ait2020kbot,chen2021intelligent} methods were developed to query KGs locally, without exposing data to third parties. These systems typically translated natural language questions into formal query languages. However, they were usually supervised and thus inherited well-known limitations: (1) the need for large training datasets, which are often unavailable; (2) poor generalization to questions that differ slightly from the training set; and (3) high adaptation costs for new use cases.

Motivated by these challenges, this paper investigates a privacy-preserving approach that leverages the semantic capabilities of LLMs without disclosing sensitive information. Given a property graph as input and without requiring additional training data, our method automatically: (1) extracts the graph schema (node labels, node properties, relation labels, and relation types); (2) builds a dictionary of sensitive values from node and relation properties (which can be manually refined by users); and (3) trains a Named Entity Recognition (NER) system to recognize the identified labels, properties, and values.

\begin{figure*}[htbp]
    \centering
    \includegraphics[width=.7\linewidth]{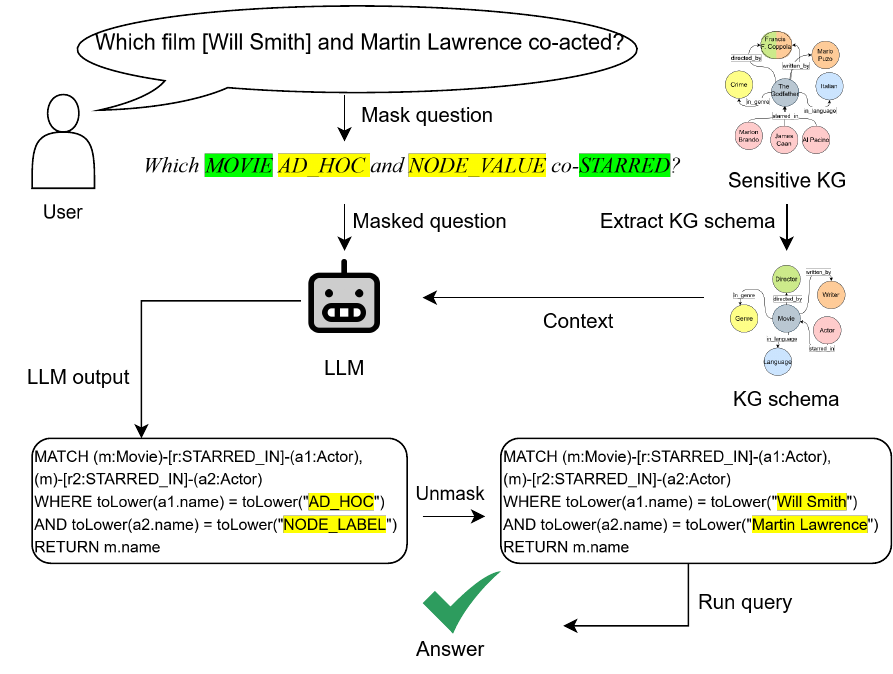}
    \caption{Workflow of the \textit{AskSafely} methodology for generating queries with LLMs without disclosing sensitive information.}
    \label{fig:asksafely}
\end{figure*}

As illustrated in Figure \ref{fig:asksafely}, natural language questions provided by the user are then reformulated to improve clarity with respect to the graph schema, while sensitive values are masked. The LLM receives this reformulated question and is asked to generate the corresponding Cypher~\cite{francis2018cypher} query based solely on the graph schema passed as context. Finally, placeholders in the generated query are replaced with the original sensitive values. An additional benefit of passing only the graph schema is that it significantly reduces the number of tokens needed, which both lowers cost and improves efficiency. Moreover, the schema can be restricted to the Role-Based Access Control (RBAC)~\cite{ferraiolo1995role} sub-schema authorized for the user, ensuring that generated queries never target parts of the graph beyond user's access rights. The resulting query can be returned to the user or executed on the graph, with the results made available accordingly.

To evaluate this privacy-aware query generation approach, we conduct an ablation study examining: (1) the effect of improving question clarity through schema information; and (2) the impact of masking sensitive data before query generation. In addition, we perform a privacy analysis to quantify both empirical and theoretical privacy gains.
\section{Related Works}
\label{sec-related_works}

We divide KG-based question answering (Q\&A) methods into four categories: (1) rule-based; (2) classical machine learning (ML); (3) transformer-based; and (4) LLM-based.

Rule-based methods rely on predefined rules to map natural language questions to KG query languages such as SPARQL and Cypher. They are deterministic and do not involve machine learning algorithms. For example, \cite{meloni2023integrating} proposes a chatbot for querying academic data, where user intent is recognized through predefined rules and managed by a finite-state system. Such methods do not require training data and provide interpretable outputs. However, they are strictly limited to the rules defined in advance, which constrains their applicability in practice.

Classical ML methods also rely on predefined rules but incorporate machine learning algorithms such as SVM~\cite{ait2020kbot}, XGBoost, and Maximum Entropy Markov Models~\cite{chen2021intelligent}. These approaches learn mappings from natural language to SPARQL or Cypher using training examples. While they outperform rule-based methods, they require the acquisition and preparation of training data, which introduces additional costs.

Transformer-based methods leverage the transformer architecture~\cite{vaswani2017attention} to improve contextualization and generalization. Those methods --- such as \cite{aghaei2022question2}, \cite{phan2025novel}, and \cite{tang2025adapting} --- typically rely on BERT variants fine-tuned on examples of expected user queries. This line of work has substantially improved the accuracy of KG Q\&A and continues to achieve state-of-the-art results~\cite{schneider2024evaluating}. However, existing approaches are usually developed for specific knowledge graphs, which limits their transferability. Applying them to new domains requires significant effort to design specialized solutions and to collect and clean large datasets of training queries.

LLM-based methods rely on generative large language models such as GPT~\cite{achiam2023gpt}, LLaMA~\cite{dubey2024llama}, and DeepSeek~\cite{liu2024deepseek}. These approaches achieve excellent performance \cite{schneider2024evaluating} thanks to their generalization and contextualization capabilities, often without requiring training data. In fact, their strongest results are obtained via zero-shot prompting~\cite{zhu2024llms}. Furthermore, recent studies~\cite{nuutila2025,mandilara2025decoding} have shown that LLMs can generate accurate queries even when provided solely with the graph schema as contextual information. However, LLMs are prone to hallucinations, which must be carefully mitigated. Moreover, due to their large number of parameters, hosting state-of-the-art LLMs locally is impractical for most users. Consequently, applications typically rely on third-party services provided by companies such as OpenAI, Google, or DeepSeek AI.

In scenarios where labeled training data is scarce or nonexistent and where sensitive data is involved, users are left with two imperfect options: rule-based systems or classical ML methods that require less training data. Yet, as noted above, both categories have limited semantic ability to interpret user questions, since anything not explicitly covered by predefined rules or training examples often produces suboptimal results. Therefore, developing an approach that enables effective Q\&A over KGs while addressing the privacy concerns of relying on third-party generative LLMs remains an open research problem. Table~\ref{tab:qa} summarizes the characteristics of existing KG-based Q\&A methods and highlights how our proposed privacy-aware LLM approach compares.

\begin{table}[htbp] 
    \centering
  \begin{tabular}{lccc} 
      \hline 
      KG-based Q\&A & \multicolumn{1}{l}{\begin{tabular}[c]{@{}l@{}}No additional\\ training/data\end{tabular}} & \multicolumn{1}{l}{Generalization} & \multicolumn{1}{l}{Privacy} \\ 
      \hline 
      Rule-based & X & & X \\ \arrayrulecolor{gray!50} \hline 
      Classical ML & & X & X \\ \hline 
      Transformers-based & & X & X \\ \hline 
      LLM-based & X & X & \\ \hline 
      \begin{tabular}[c]{@{}l@{}}\textbf{Privacy-aware LLM}\\ \arrayrulecolor{black}\textbf{(our method)}\end{tabular} & \textbf{X} & \textbf{X} & \textbf{X} 
  \end{tabular} 
  \caption{Comparison of KG-based Q\&A methods, with ``X'' marking characteristics those methods have.} 
  \label{tab:qa}
\end{table}
\section{Privacy-Aware Q\&A}
\label{sec-methodology}

In this paper, we investigate how to leverage the contextualization and generalization capabilities of LLMs to automatically query sensitive data from KGs using natural language questions, which may themselves contain sensitive information. We propose \textit{AskSafely}, a privacy-aware query generation method that enables users to rely on third-party generative LLM services without sharing sensitive data --- critical in scenarios constrained by organizational policies, ethical standards, non-disclosure agreements, or legal frameworks. Throughout this paper, we illustrate our method using the METAQA dataset~\cite{zhang2018variational}, which, although it does not contain sensitive data, is treated as if all values in the KG were sensitive.

When generating KG queries from natural language questions using LLMs, there are two main sources of potential sensitive data leakage. First, the contextual information provided to the LLM, which often includes the data from the KG itself~\cite{avila2024framework,alekseev2025benefits,tian2025systematic,ao2025lightprof,ma2025debate}. Second, the user’s question may contain sensitive information in its text; for example ``Other than Bad Boys, which movies did Will Smith and Martin Lawrence co-star in?'' implies prior collaboration between the actors, which could be considered private information.

Below, we describe how our method addresses each of these challenges.

\subsection{Graph schema as context}

\begin{figure}[h]
\centering
\begin{subfigure}{0.49\linewidth}
\centering
\includegraphics[width=\linewidth,trim=7mm 7mm 6mm 7mm,clip]{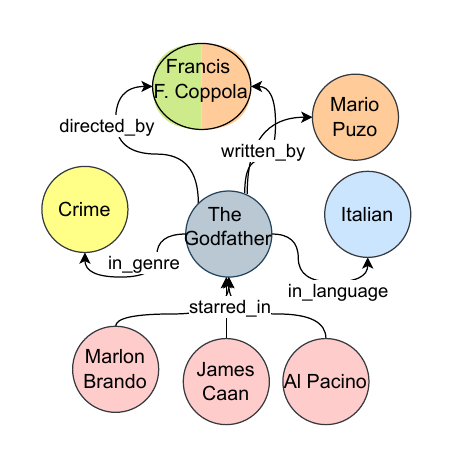}
\caption{Sensitive KG.}
\label{fig:kg}
\end{subfigure}
\hfill
\begin{subfigure}{0.49\linewidth}
\centering
\includegraphics[width=\linewidth,trim=7mm 7mm 7mm 7mm,clip]{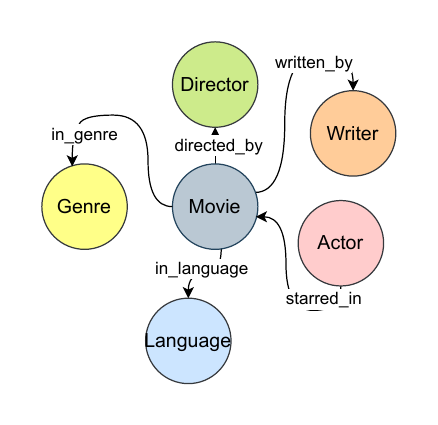}
\caption{KG schema.}
\label{fig:structure}
\end{subfigure}
\caption{``The Godfather'' sub-KG and its schema.}
\label{fig:kg_and_structure}
\end{figure}

Our method relies on the observation that KGs used for Q\&A tasks are typically based on an ontology or predefined schema. Figure~\ref{fig:kg_and_structure} illustrates this concept.

Figure~\ref{fig:kg} shows a KG containing seven nodes connected to the movie ``The Godfather''. This graph is constructed based on the schema in Figure~\ref{fig:structure}, which includes six node types: ``Movie'', ``Writer'', ``Director'', ``Genre'', ``Language'', and ``Actor''. The schema dictates node types, their attributes, and relationships. While the data stored in the KG may be sensitive, the schema itself is not, and can therefore provide sufficient context for the LLM to generate executable queries without accessing sensitive values. In addition to preserving privacy, providing only the graph schema substantially reduces the prompt size. This improves token efficiency, which directly translates into lower costs and faster query generation.

Moreover, because the schema can be restricted to the sub-schema that the user is authorized to access, the LLM is prevented from generating queries that target parts of the KG beyond their access rights. This integrates access control directly into the query-generation pipeline, ensuring that privacy is preserved not only at the level of sensitive values, but also at the level of graph topology.

In detail, we represent the schema of a knowledge graph using its fundamental components:  

\begin{itemize}
    \item Node labels (\texttt{NODE\_LABEL}), which define the entity types represented in the graph;
    \item Node properties (\texttt{NODE\_PROPERTY}) and node values \\(\texttt{NODE\_VALUE}), describing the properties of nodes and their corresponding instances;  
    \item Relation labels (\texttt{RELATION\_LABEL}), which specify the types of edges linking nodes;
    \item Relation properties (\texttt{RELATION\_PROPERTY}) and relation values (\texttt{RELATION\_VALUE}), describing the properties of relations and their corresponding instances;
\end{itemize}

As illustrated in Figure~\ref{fig:kg_and_structure}, the label “Director” is a NODE\_LABEL, one could infer the existence of ``name'' as a NODE\_PROPERTY, and ``Francis F. Coppola'' corresponds to a NODE\_VALUE. The relation ``directed\_by'' represents a RELATION\_LABEL connecting the movie node to the director node. In this example, relations have no additional properties or values, but if present, they would follow the same representation pattern as those defined for nodes. 

\subsection{Masking user questions}
\label{sec:mask-question}

User questions may contain sensitive information and domain-specific vocabulary that is difficult for the LLM to interpret when the context given omits private information. To address this, we identify two sets of entities: (i) sensitive entities that must not be sent directly to the LLM, and (ii) key entities critical for query generation, representing node labels, relations, and attributes.

For example, consider the question below, which have sensitive data explicitly flagged between brackets.

\vspace{0.5em}
\noindent - \textbf{Sensitive question:} \\\textit{Which film [Will Smith] and Martin Lawrence co-acted?}
\vspace{0.5em}

Based on the graph schema shown in Figure~\ref{fig:structure}, we identify the following fundamental components of the KG as entities: \\NODE\_LABEL, NODE\_PROPERTY, NODE\_VALUE, RELATION\_LABEL, RELATION\_PROPERTY, and RELATION\_VALUE.
Additionally, we introduce the AD\_HOC entity type, which represents user-specified sensitive values not explicitly present in the KG.
These entities are then masked to ensure that no sensitive information is sent to the LLM.

\vspace{0.5em}
\noindent - \textbf{Pre-processed question:}\\ \textit{Which NODE\_LABEL AD\_HOC and NODE\_VALUE co-RELATION\_LABEL?}
\vspace{0.5em}

Next, key entities are substituted with their canonical KG synonyms to reduce the risk of the LLM using incorrect terminology. For instance, ``film'' is replaced by ``MOVIE'' and ``acted'' by ``STARRED'', matching the KG labels and properties. The resulting privacy-aware question submitted to the LLM is:

\vspace{0.5em}
\noindent - \textbf{Privacy-aware question:} \\\textit{Which MOVIE AD\_HOC and NODE\_VALUE co-STARRED\_IN?}
\vspace{0.5em}

Key and sensitive entities can be identified using a NER algorithm appropriate to the user’s domain. In our implementation, this step is performed using a simple rule‑based approach with case‑insensitive matching between question tokens and KG entities, including their manually defined synonyms. More advanced NER models could also be used in practice.

\subsection{Privacy-aware LLM prompt and reply}

Once the KG context and user question are free from sensitive information, we construct a zero-shot prompt instructing the LLM to generate a Cypher query to retrieve the requested information. Because the prompt includes only schema-level information rather than the entire graph, it remains compact and token-efficient, allowing us to scale to larger KGs without overwhelming the LLM input. The prompt we use throughout our experiments can be visualized in Appendix \ref{sec:main_prompt}.

As an example, the LLM’s response to the privacy-aware question described in Section~\ref{sec:mask-question} is shown below:

\vspace{0.5em}
\noindent - \textbf{LLM reply:}
\begin{lstlisting}
MATCH (m:Movie)-[r:STARRED_IN]-(a1:Actor),
(m)-[r2:STARRED_IN]-(a2:Actor)
WHERE toLower(a1.name) = toLower("AD_HOC")
AND toLower(a2.name) = toLower("NODE_LABEL")
RETURN m.name
\end{lstlisting}

Before execution, placeholders are replaced with the original sensitive values provided by the user, yielding the final query:

\vspace{0.5em}
\noindent - \textbf{Final Cypher query:}
\begin{lstlisting}
MATCH (m:Movie)-[r:STARRED_IN]-(a1:Actor),
(m)-[r2:STARRED_IN]-(a2:Actor)
WHERE toLower(a1.name) = toLower("Will Smith")
AND toLower(a2.name) = toLower("Martin Lawrence")
RETURN m.name
\end{lstlisting}

Finally, the Cypher query is executed, and the results returned.
\section{Experiments \& Results}
\label{sec-experiments}

Considering that the capability of LLMs to accurately generate Cypher queries using the graph schema has already been demonstrated~\cite{nuutila2025,mandilara2025decoding}, our objective is twofold: (1) to investigate the impact of applying the \textit{AskSafely} privacy constraints on the quality of the queries generated by these models, and (2) to assess both the practical and theoretical privacy gains achieved through the use of \textit{AskSafely}.

To implement the privacy-aware strategy for querying KGs described in Section~\ref{sec-methodology}, we used the BESSER Agentic Framework (BAF)~\cite{BAF}. BAF is designed to facilitate and accelerate the development of AI agents in a low-code manner. It natively implements the simple NER method described previously and allows the definition of synonyms, which are automatically used to identify key entities. Additionally, BAF integrates generative-LLM APIs, enabling seamless use of LLMs during agent execution. All the code used in our experiments is available on GitHub\footnote{\url{https://github.com/maurodlt/PrivateNL2CYPHER}}.

\smallskip
\textbf{Dataset}: We evaluate our strategy using the METAQA dataset~\cite{zhang2018variational}, which contains facts for 16,427 movies, including their actors, directors, writers, release years, language, tags, genre, IMDB votes, and IMDB ratings. The dataset also provides natural language questions and corresponding answers, with entities already annotated between brackets. We use the ``vanilla'' test set, containing 39,093 questions. Most of these questions follow the same pattern, varying only in entity values (e.g., actor or movie names). Because these values are automatically replaced in our privacy-aware strategy, we reduced the dataset to 503 questions with unique patterns. A k-hop query refers to a question that requires traversing k relations in the knowledge graph to reach the answer. The METAQA dataset includes questions requiring one-, two-, and three-hop reasoning, making it a challenging benchmark that covers both simple and complex reasoning tasks. Among the 503 unique questions, 143 require 1-hop, 210 require 2-hop, and 150 require 3-hop queries to correctly answer the question.

\subsection{Experimental setting}

First, we created a property graph in Neo4J~\cite{neo4j2025} using the triples from METAQA. The graph follows the schema illustrated in Figure~\ref{fig:structure}. Each node has a ``name'' property, and ``Movie'' nodes also include properties corresponding to their connected entities and movie-specific attributes: ``release year'', ``tags'', ``IMDB votes'', and ``IMDB ratings''.

Next, we extract from the graph and provide to BAF the node types, relations, properties, property values, and their synonyms. 
In our implementation, most steps of the pipeline are automatic. The graph schema and property values are automatically extracted from Neo4J, and masking, synonym substitution, prompt construction, query generation, and query execution are performed automatically by BAF and the LLM. The only manual configuration concerns the definition of synonyms and the specification of which entity values are considered sensitive --- when all values in the graph are treated as sensitive, this configuration step can also be automated.

For illustration, we manually defined a few synonyms, shown in Table~\ref{tab:synonyms}. In practice, most synonyms for non-sensitive entities (e.g., node or relation labels) can be automatically suggested by the LLM itself, as they do not contain private information.
In contrast, sensitive entities -- such as property values corresponding to personal names or other private data -- require special attention. Synonyms for these entities should not be generated by the LLM but instead obtained from a controlled external list or through a domain-adapted NER algorithm, depending on the specific use case.
Nonetheless, such synonyms are primarily used to enhance the entity recognition process and are not strictly required for our approach to function effectively.

\begin{table}[htp]
    \centering
    \begin{tabular}{l|l}
         \textbf{Entity} & \textbf{Synonyms} \\\hline
         Movie & film, movie, films, movies\\
         directed\_by & directed \\
         in\_language & language \\
         release\_year & year, release, released \\
         has\_tags & tags, tag, described, about \\
         written\_by & wrote, written \\
         starred\_actors & actor, actress, actors, actressess, star, starred \\
         has\_imdb\_votes & votes, vote \\
         has\_genre & genre, type \\
         has\_imdb\_rating & rating \\
         name & names
    \end{tabular}
    \caption{List of synonyms defined in BAF.}
    \label{tab:synonyms}
\end{table}

We then define which entities are sensitive. For this experiment, all property values in the KG are considered sensitive, except for single-word movie titles and movie tags. Terms like ``Creator'', ``Fall'', and ``Primary'' (movie titles) or ``main'' and ``script'' (tags) are not considered sensitive, as they are unlikely to reveal private information and are more likely to appear as common words in contexts unrelated to movies or tags.

\subsection{Q\&A Performance}

We evaluated our agent on the 503 selected questions and compared the generated query results with the corresponding dataset answers. Table~\ref{tab:results} presents the accuracy obtained under three configurations: (1) GPT‑based query generation without privacy constraints, (2) our privacy‑aware approach without synonym substitution, and (3) our full privacy‑aware approach with synonym substitution. To better understand how question complexity affects performance, results are additionally separated according to the number of hops (1‑hop, 2‑hop, and 3‑hop) required to answer each question. For each configuration, both the automatically computed accuracy and the manually verified accuracy are reported. The manual evaluation was necessary because the dataset’s reference answers were sometimes incomplete. For instance, for the question about the genre of the movie ``Bad Boys'', the dataset expects [``Drama'', ``Comedy''], even though the underlying knowledge graph includes additional genres such as [``Drama'', ``Comedy'', ``Crime'', ``Action''].

\begin{table*}[htp]
\centering
\setlength{\tabcolsep}{5pt}

\begin{tabular}{lcccccccc}
\toprule
\textbf{Method} &
\multicolumn{2}{c}{\textbf{1-Hop}} &
\multicolumn{2}{c}{\textbf{2-Hop}} &
\multicolumn{2}{c}{\textbf{3-Hop}} &
\multicolumn{2}{c}{\textbf{Overall}} \\

\cmidrule(lr){2-3}
\cmidrule(lr){4-5}
\cmidrule(lr){6-7}
\cmidrule(lr){8-9}

& Acc & M-Acc & Acc & M-Acc & Acc & M-Acc & Acc & M-Acc \\
\midrule

\makecell[l]{GPT-5 \\ (no privacy)}
& 81.0 & 90.1 & 92.9 & 97.7 & 74.0 & 92.7 & 83.9 & \textbf{94.0} \\[10pt]

\makecell[l]{Privacy-aware \\ (no synonym)}
& 85.9 & 95.1 & 78.6 & 82.9 & 55.3 & 62.7 & 73.7 & \textbf{80.3} \\[10pt]

\makecell[l]{Privacy-aware \\ (complete)}
& 91.5 & 97.2 & 85.7 & 89.6 & 63.3 & 73.3 & 80.7 & \textbf{86.9} \\

\bottomrule

\end{tabular}
\caption{Accuracy of GPT‑5 with and without the proposed privacy‑aware methodology, separated by the number of hops required to answer each question. \textit{Acc} denotes automatically computed accuracy, while \textit{M‑Acc} denotes manually verified accuracy. The final columns report overall results across all questions.}
\label{tab:results}
\end{table*}

By substituting the sensitive tokens present in the users' questions by placeholders, we avoid leakage of sensitive data to third party companies. Still, to guarantee the increase in privacy in those setting we perform 3 analysis on the privacy gained when using \textit{AskSafely}. First, we track if and how many sensitive tokens were exposed during our experiments. Second, we test the possibility of indirect leakage of sensitive information. Third, we calculate the reduction of mutual information between the users question and sensitive tokens before and after the masking.

\subsubsection{Token Exposure}

We calculate the total number of sensitive tokens exposed during the executing of \textit{AskSafely} in the MetaQA dataset. Considering that the dataset already flags sensitive tokens, as expected, \textit{AskSafely} had 0 tokens exposed during the experiments.   

\subsubsection{Leakage Test}
\label{sec:leak}

We empirically evaluate if it was possible to infer sensitive tokes from the remaining context passed to the LLMs. For this, we asked \textit{gpt5-mini} to guess the masked values based on the context previously given to the LLM via the prompt in Appendix \ref{sec:leakage_prompt}.

After manually comparing the LLM guesses with the sensitive values contained in the original users' questions, we observed that the LLM could not make a single correct guess. In general, the LLM mostly just guessed the name of famous actors such as Tom Hanks, famous directors such as Christopher Nolan, and famous movies such as The Matrix, and Inception.

\subsubsection{Mutual Information Analysis}

Beyond the empirical leakage test, we further estimate how much information about the sensitive values may still be inferred from the transmitted data. To this end, we interpret mutual information $I(X;Z)$ as a measure of how much dependence remains between sensitive inputs and observable outputs~\cite{unsal2023information}, considering the sensitive entities $X$ and the users’ questions, both in their masked form $Z$ and in their original form $Z'$:

\begin{equation}
    I(X;Z) = \sum_{z \in Z}\sum_{x \in X} P_{(X,Z)}(x,z)\log\left(\frac{P_{(X,Z)}(x,z)}{P_X(x)P_Z(z)}\right).
\end{equation}

Since we work with textual data, estimating exact probability distributions is not feasible. Instead, we approximate them using the distribution of embedding similarities. Specifically, we derive pseudo-probabilities from cosine similarities between the embeddings of $X$ and $Z$, normalized through a softmax transformation:
\begin{equation}
    I(X;Z)
    \approx
    \sum_{i,j}
    \tilde{p}(x_i,z_j)\,
    \log\!\left(\frac{\tilde{p}(x_i,z_j)}{\tilde{p}_X(x_i)\tilde{p}_Z(z_j)}\right),
    \quad
\end{equation}

\begin{equation*}
    \text{with } \tilde{p}(x_i,z_j) =
    \frac{\exp(\cos(f(x_i),f(z_j)))}%
         {\sum_{k,l}\exp(\cos(f(x_k),f(z_l)))}.
\end{equation*}

Finally, we quantify the privacy improvement achieved by masking through the relative reduction in mutual information:
\begin{equation}
    \text{Privacy gain} =
    \frac{I(X,Z) - I(X,Z')}{I(X,Z)} \times 100\%
    = \frac{0.0024 - 0.0007}{0.0024}
    = 71.1\%.
\end{equation}

Despite the already low initial mutual information between user questions and sensitive values, these results show that \textit{AskSafely} further reduces potential information leakage by approximately 71\%.

\section{Discussion}
\label{sec:discussion}

In this section, we discuss the key insights gained from the experiments described in Section~\ref{sec-experiments}.

\begin{itemize}
    \item \textbf{Privacy-awareness impact on performance.}
    Table~\ref{tab:results} shows that GPT maintains consistently high accuracy across questions of different complexity levels. In the privacy‑aware setting, accuracy decreases as the number of hops increases, which we attribute to the reduced contextual information available to the LLM after masking sensitive tokens. Nevertheless, synonym substitution recovers a substantial portion of this loss and improves performance across all hop levels. Interestingly, in the case of simple 1‑hop questions, the privacy‑aware approach with synonym substitution even slightly improves performance compared to the baseline, suggesting that aligning user vocabulary with the schema can help guide the query generation process.

    \item \textbf{Privacy and scalability are not trade-offs.} By transmitting only the graph schema, our approach not only protects sensitive information but also keeps the LLM prompt compact. In our experiments, each LLM query required fewer than 1,400 tokens in total, with less than 500 tokens used to represent the METAQA schema. Based on this efficiency, we estimate that \textit{AskSafely} can scale to knowledge graphs containing approximately $800\times$ more node types and relations than those evaluated in Section~\ref{sec-experiments} -- that is, around 4,800 node types and 4,000 relation types. It is important to note that this estimate concerns the \textit{schema} of the graph, not its instance size; in principle, \textit{AskSafely} imposes no limit on the number of nodes or edges in the underlying KG. These findings demonstrate that privacy preservation and computational efficiency are not competing objectives but rather complementary ones.

    \item \textbf{Choosing synonyms.} The importance of defining synonyms can be seen in Table \ref{tab:results}. In our experiments, most of the accuracy gain when substituting synonyms came from replacing the word ``about'' with ``has\_tags.'' While this substitution may sound unnatural in natural language, it matched the way the dataset was structured (e.g., mapping ``what topics is [Free Willy] about'' to movie tags). In practice, for questions following a pattern closer to natural language than the one provided by the MetaQA dataset, we expect synonyms automatically identified by an LLM to perform satisfactorily. Still, if a background ontology of the domain in question is available, their pre-defined synonyms could further improve \textit{AskSafely} accuracy.

    \item \textbf{LLMs also make mistakes.} The most frequent issue observed was confusion between entity labels. For example, the question ``who is listed as director of the movies starred by [Solo] actors’’ was misinterpreted, with the LLM treating ``Solo’’ as the name of an actor rather than the name of a movie, leading to incorrect results. Such error was more present in \textit{AskSafely} than pure GPT-5 due to the reduction of semantic information given to the LLM, which depending on the way the question was formulated, did not suffice for it to produce an accurate query. Another source of errors was returning more information than requested. For instance, when asked ``when were the films directed by [Peter Lord] released'', the system returned [‘Chicken Run’, 2000]’’ instead of the expected ``[2000]''. 
    
    \item \textbf{Supervised Methods and Generalization Challenges.} \cite{madani2023answering} report 100\% accuracy on the MetaQA dataset by training a sequence-to-sequence transformer model to map natural language questions into graph paths. These paths are output as Prolog functions, which can then be executed to retrieve the correct answers. However, this approach is entirely dependent on training data and lacks generalization. Small variations in phrasing that were not seen during training consistently failed. For example, questions such as ``Which movies starred [Al Pacino]?'', ``Which movies had [Al Pacino] as an actor?'', ``Which movies featured [Al Pacino]?'', or ``Which movies have [Al Pacino]?'' all produced incorrect results. After several attempts --- with the query ``Which movies [Al Pacino] starred in?’’ --- the model was able to generate the correct answer. Thus, we emphasize the need to investigate the generalization capabilities of supervised methods, especially in Q\&A settings where users are unlikely to phrase their questions according to a predefined schema.
\end{itemize}

\subsection{Limitations}

Below we highlight the following limitations we observed while developing and testing \textit{AskSafely}.

\begin{enumerate}
    \item \textbf{Not suitable for anonymous graph schemas.} If the schema of the graph contains sensitive information, such as a trade secretes, \textit{AskSafely} cannot be used by design. 
    \item \textbf{Indirect leakage of masked terms.} Despite the privacy analysis presented in Section~\ref{sec:leak}, we cannot fully guarantee that no sensitive information can be inferred from masked questions. Such inference largely depends on how the question is formulated and on the contextual information available.
    \item \textbf{Domain dependent NER.} The choice of the NER algorithm depends on the domain, and direct word matching should be avoided in practice.
    \item \textbf{Synonyms definition.} Synonyms for sensitive values must be defined according to the specific use case, which may be impractical for large knowledge graphs.
\end{enumerate}

Nevertheless, these limitations are unlikely to affect the applicability of \textit{AskSafely} in most scenarios. This claim is supported by the following observations:

\begin{itemize}
    \item Graph schemas seldom contain sensitive information;
    \item The leakage tests can be easily adapted to different domains using representative samples of expected question types;
    \item The same NER tools and thesauri or ontologies used to build or maintain the knowledge graph can also be leveraged by \textit{AskSafely}. When domain-specific resources are unavailable, generic ones may still yield satisfactory results, as demonstrated in Section~\ref{sec-experiments}.
\end{itemize}

\section{Conclusion}
\label{sec-conclusion}
We introduced a privacy-aware strategy for generating Cypher queries with third-party generative LLMs. By using the graph schema as context and masking sensitive information in user questions, our approach preserves privacy while still benefiting from the strong generalization capabilities of LLMs. Moreover, relying only on the graph schema keeps prompts compact and token-efficient, enabling scalability to larger graphs. Experiments on the METAQA dataset achieved up to 86.9\% accuracy with 0\% direct sensitive information leakage and 71\% reduction in potential indirect leakage, demonstrating that privacy can be protected while maintaining query quality. In future work, we aim to extend this approach to domain-specific use cases. We also plan to automatically filter and adapt the graph schema to the RBAC model provided for the users. Together, these directions will allow us to explore how privacy-preserving and access control mechanisms can be integrated with LLM-based reasoning at scale.

\begin{acks}
This project is supported by the Luxembourg National Research Fund (FNR) PEARL program, grant agreement 16544475.
\end{acks}

\bibliographystyle{ACM-Reference-Format}
\bibliography{sample-base}

\appendix
\renewcommand{\thesubsection}{\arabic{subsection}}

\clearpage
\section*{\appendixname}
\label{sec:appendix}

\subsection{\textit{AskSafely} Main prompt}
\label{sec:main_prompt}
{\footnotesize
\begin{verbatim}
You are a tool that transforms natural language questions into Cypher queries.  
    The graph structure is: {graph_structure}.  
    
    Your task: Generate a single Cypher query that best represents the question: 
    {masked_message}.  
    
    STRICT RULES (must always be followed):
    1. Output **only** the Cypher query (no explanations, no extra text, 
    no formatting).  
    2. Never include relationship types or directions.  
       - Always write relationships as `-[r]-`.  
       - Forbidden: `-[r]->`, `<-[r]-`, `[:RELATION_TYPE]`.  
       - Wrong: `(a)-[r]->(b)` → Correct: `(a)-[r]-(b)`  
    3. Always restrict nodes by labels **as they appear in `graph_structure`**.  
       - If a node label exists in `graph_structure`, you must include it 
       (e.g. `(m:Movie)`, `(w:Writer)`).  
       - Do not invent or guess labels.  
       - Do not omit labels when they are defined in `graph_structure`.  
    4. Never return whole nodes. Always return a property and remove duplicates:  
       - Example: `RETURN DISTINCT w.name`  
    5. Example of bad vs good:  
    - Bad (labels missing): MATCH (m)-[r]-(w) WHERE m.name = 'AD_HOC' 
    RETURN DISTINCT w.name
    - Good (labels enforced from graph structure): MATCH (m:Movie)-[r]-(w:Writer) 
    WHERE toLower(m.name) = toLower('AD_HOC') RETURN DISTINCT w.name  

    Failure to follow these rules will result in an invalid answer.    
\end{verbatim}
}

\subsection{Leakage Test Prompt}
\label{sec:leakage_prompt}

{\footnotesize
\begin{verbatim}
You are an assistant specialized in reconstructing anonymized elements 
using structural graph context.
The question you will see contains one placeholder: AD_HOC.
The only available contextual information comes from graph_structure,
which describes the relationships or topology relevant to the question.
Your task: use graph_structure to infer the most plausible replacement
for AD_HOC.
Output strictly: return the inferred replacement text only — no commentary, 
formatting, or punctuation.

Question: {masked_question} 

graph_structure: {graph_structure}

Reply only with your guess, do not explain.
\end{verbatim}
}

\end{document}